\documentclass{article}
\usepackage{amssymb}
\usepackage{amsmath}
\usepackage{xcolor} 

\begin{document}

\def\pod#1#2{\mathop{#1}_{#2}}
\def\zvez#1#2{\mathop{#1}^{#2}{}}

\newtheorem{theo}{Theorem}
\newtheorem{defi}{Definition}
\newtheorem{Opredelenie}{Definition}

\title{SHAPOVALOV WAVE-LIKE SPACETIMES
}

\author{
Konstantin~Osetrin\thanks{osetrin@tspu.edu.ru},  \\[1ex]
Evgeny~Osetrin\thanks{evgeny.osetrin@gmail.com}, \\[2ex]
Tomsk State Pedagogical University, \\[1ex]
Tomsk, 634061, Russia\\
}

\date{July 22, 2020}

\maketitle

\begin{abstract}
A complete classification of space-time models is presented, which admit the privileged coordinate systems, where the Hamilton-Jacobi equation for a test particle is integrated by the method of complete separation of variables with separation of the isotropic (wave) variable, on which the metric of space depends (wave-like Shapovalov spaces). For all types of Shapovalov spaces, exact solutions of the Einstein equations with a cosmological constant in vacuum are found. Complete integrals are presented for the eikonal equation and the Hamilton-Jacobi equation of motion of test particles.
\end{abstract}

\section*{Introduction}

In this paper, we consider space-time models that allow integration by the method of complete separation of variables of the eikonal equation and the Hamilton-Jacobi equation for the motion of test particles in a gravitational field,
which allow the ''privileged'' coordinate systems (where separation of variables is possible), in which one of the variables is isotropic (wave).

Spaces that allow complete separation of variables in the Hamilton-Jacobi equation for test particles are called St{\"{a}}ckel spaces after Paul St{\"{a}}ckel, who first posed this problem (Paul St{\"{a}}ckel, see \cite{Stackel1897145}). In a series of papers, St{\"{a}}ckel solved this problem for the case when the metric of space in a privileged coordinate system has a diagonal form.
The theory of St{\"{a}}ckel spaces was developed by many authors and was finally completed in the period 1973-1980 in the works of V.N.~Shapovalov, who was the first to build a complete classification of these spaces and obtain a general form of their metrics in privileged coordinate systems, where complete separation of variables is allowed (see \cite{Bagrov19731533, Naiminov1974571, Shapovalov19751650, Shapovalov19781124, Shapovalov19781130, Shapovalov1978718, Shapovalov1979790}).
We recall some details from the theory of St{\"{a}}ckel spaces (for a more complete presentation, see f.e.
\cite{Shapovalov1979790, Obukhov2004127, Benenti2016}).

Let us consider the Hamilton-Jacobi equation for the motion of a test particle of mass $ m $ in a gravitational field defined by the metric tensor $ g_{ij} $ in the coordinate system $ \{ x^i \} $:
\begin{equation}
g^{ij}\frac{\partial S}{\partial x^i}\frac{\partial S}{\partial x^j}=m^2,
\qquad
i,j,k=1...n,
\label{HJE}
\end{equation}
where the capital letter $ S $ denotes the action function of a test particle, $ n $ is the dimension of space.
To avoid confusion, note that in what follows we will also use the lowercase $ s $ to denote a space-time interval.
\begin{Opredelenie}
\label{Stackel_space}
If the space admits the existence of a {\bf ''privileged''} coordinate system $ \{x^i \} $, where the Hamilton-Jacobi equation 
(\ref{HJE}) admits complete separation of variables, when the complete integral for the action function of the test particle $ S $ can be written as:
\begin{equation}
S=\phi_1(x^1,\lambda_1,...,\lambda_n)+\phi_2(x^2,\lambda_1,...,\lambda_n)+...+\phi_n(x^n,\lambda_1,...,\lambda_n)
,
\end{equation}
$$
\lambda_1,...,\lambda_n-\mbox{const},
$$
$$
\det \left|\frac{\partial{}^2 S}{\partial x^i \partial \lambda_j}\right| \ne 0
,
$$
then such a space is called\, {\bf St{\"{a}}ckel space}, and the parameters $ \lambda_i $ are the
constants of separation.
\end{Opredelenie}
\begin{Opredelenie}
\label{Conformal_Stackel_space}
Space with metric $ \tilde g_{ij} $, allowing complete separation of variables in the eikonal equation for radiation propagation (where $\Psi $ is the eikonal function):
\begin{equation}
\tilde g^{ij}\Psi_{,i}\Psi_{,j}=0
,
\label{Eikonal}
\end{equation}
is called a {\bf conformal St{\"{a}}ckel space}.
\end{Opredelenie}
Conformal St{\"{a}}ckel spaces admit separation of variables in the same privileged coordinate system as a St{\"{a}}ckel spaces, and the metrics of such spaces, as can be seen, differ from the metrics of St{\"{a}}ckel spaces by an arbitrary conformal factor.

The following theorem was proved by V.N. Shapovalov (see \cite{Shapovalov19781124,Shapovalov1979790}):
\begin{theo}
Let $ V_n $ be a St{\"{a}}ckel space. Then the components of the metric tensor $ g_{ij} $
in a privileged coordinate system can be written as
\begin{equation}
g^{ij}=(\Phi^{-1})_n{}^\nu \,  G^{ij}_\nu,
\qquad
G^{ij}_\nu=G^{ij}_\nu(u^\nu), 
\qquad
\Phi_\mu{}^\nu=\Phi_\mu{}^\nu(u^\mu), 
\end{equation}
\begin{equation}
G^{ij}_\nu(u^\nu)=
\delta^i_\nu\delta^j_\nu \, \varepsilon_\nu(u^\nu) +
(\delta^i_\nu\delta^j_p+\delta^j_\nu\delta^i_p) \, G^{\nu p}_\nu(u^\nu) +
\delta^i_p\delta^j_q\,G^{pq}_\nu(u^\nu),
\quad
\label{nosum}
\end{equation}
$$
\mbox{(there is no summation over index $\nu$ in (\ref{nosum}) )},
$$
$$
p,q=1,...N,
\quad
\nu,\mu=N+1,...n,
$$
where  $\Phi_\mu{}^\nu(u^\mu)$  is the so-called St{\"{a}}ckel matrix.
\end{theo}
Summation over repeated superscripts and subscripts is accepted, except for the expression (\ref{nosum}).
The $ N $ is the number of "ignored"\,  variables that the metric in the privileged coordinate system does not depend on. The subscripts $ p, q $ number "ignored"\,  variables, and the indices $ \nu, \mu $ number "nonignored"\, variables.

It is shown that the equation of geodesics in St{\"{a}}ckel
spaces admits first integrals that commute pairwise with respect to
Poisson brackets
\begin{equation}
\pod X\mu=(\Phi^{-1})^\nu_\mu H_\nu, 
\qquad
H_\nu=\varepsilon_\nu p^2_\nu + 2G^{\nu p}_\nu p_pp_\nu +
h^{pq}_\nu p_pp_q,
\label{IntegralsHJE2}
\end{equation}
\begin{equation}
\pod Yp=\pod Yp{}^ip_i.  
\label{IntegralsHJE1}
\end{equation}

Thus, for the covariant characteristic
of the St{\"{a}}ckel space, it suffices to find the corresponding properties of the integrals
(\ref{IntegralsHJE2}) and (\ref{IntegralsHJE1})
in an arbitrary coordinate system
$\{x^i\}_n$. 

Let us write the functions
 $\pod X\nu$, $\pod Yp$ 
 in the form
\begin{equation}
\pod X\nu=\pod X\nu {}^{ij}p_ip_j,\qquad \pod Yp=\pod Yp{}^ip_i.
\label{KillingFields12}
\end{equation}
Then for 
$\pod X\nu {}^{ij}$ and $\pod Yp{}^i$ 
we get
\begin{equation}
\pod X\nu{}_{(ij;k)}=\pod Yp{}_{(i;j)}=0,
\end{equation}
where semicolon means covariant derivative and parentheses mean
symmetrization.

Therefore, 
$\pod Yp{}^i$, $\pod X\nu{}^{ij}$ 
are the components of the Killing vector and Killing tensor fields, respectively.

\begin{defi}
Pairwise commuting Killing vector fields
$\pod Yp{}^i$, 
where $ p = 1, ... N $ and the Killing tensor fields of the second rank
$\pod X\nu{}^{ij}$, где $\nu=N+1,...n$
form a {\bf complete set} of type $ (N.N_0) $ if the following conditions are met:
\begin{equation}
B^{pq}\pod Yp{}^i\pod Yq{}^j + B^\nu\pod X\nu{}^{ij}=0 \quad
\Longrightarrow \quad B^{pq}=B^\nu=0, 
\end{equation}
\begin{equation}
rank||\,\pod Yp{}^i\pod Yq{}_i\,||=N-N_0,   
\end{equation}
\begin{equation}
\pod X\nu{}^{ik}\pod X\mu{}^j{}_k=C^{pq}_{\nu\mu} \pod Yp{}^i\pod
Yp{}^j + C^\tau_{\nu\mu}\pod X\tau{}^{ij}, 
\end{equation}
\begin{equation}
\pod X\nu{}^{ij}\pod Yp{}_j=C^q_{\nu p}\pod Yq{}^i.
\end{equation}
\end{defi}
\begin{theo}
A necessary and sufficient geometric criterion for the St{\"{a}}ckel space
is the existence of a complete set of type
($N.N_0$).
\end{theo}
This theorem was proved by V.N. Shapovalov in the work \cite{Shapovalov1979790}.
Thus, the Hamilton-Jacobi equation can be integrated by
the method of complete separation of variables if and only if there is
a complete set of first integrals of motion.

You can define the type of St{\"{a}}ckel space as follows:
\begin{defi}
Space-time is called a St{\"{a}}ckel space of the type
($ N.N_0 $)
if there is a complete set of type ($ N.N_0 $).
\end{defi}
All these theorems and definitions are valid if
the free Hamilton-Jacobi equation for the motion of test particles in a gravitational field is considered.

St{\"{a}}ckel spaces are determined by the presence of the so-called ''complete set'' of commuting Killing fields of the first and second rank, corresponding to an additional set of algebraic requirements.
Moreover, the $ N $ commuting Killing vectors (where $ 0 \le N \le n $) included in the  ''complete set'' determine the choice of a ''privileged'' coordinate system where separation of variables is allowed and where the metric does not depend on the corresponding $ N $ variables.
\begin{Opredelenie}
\label{Ignorable}
Coordinate variables of the privileged coordinate system, on which the space-time metric does not depend, are called\, {\bf ignored}.
\end{Opredelenie}
St{\"{a}}ckel and conformal  St{\"{a}}ckel spaces are of great interest for metric theories of gravity, since they allow explicitly, in quadratures, to integrate the equations for the motion of test particles and for radiation propagation and, thereby, determine the form of geodesic lines of space-time along which test particles move in gravitational field. The application of these mathematical tools is possible in various metric theories of gravity, including modified theories of gravity with various types of matter
(see \cite{Bagrov19881141,Osetrin20066641,Obukhov200242,Makarenko1999889,Bagrov1997995,Bagrov1996744,Bagrov19981149,Osetrin20181383,Osetrin2019292}).

\section{Shapovalov spaces}

The space of the orbits of the $ N $ - parametric Abelian group of motions of the St{\"{a}}ckel space defined by the Killing vectors from the ''complete set'' can be ''isotropic'' in the sense that the restriction of the metric to these orbits can
have a determinant equal to zero. Such spaces were first found and classified by V.N.~Shapovalov, and he called them {\bf isotropic St{\"{a}}ckel spaces}.

If the Killing vectors are from the ''complete set''
$ Y^ i_ {(p)} $, where $ p $ numbers the Killing vectors in the set ($ p, q = 1, ..., N $), then for isotropic St{\"{a}}ckel spaces
in the case of four-dimensional space-time, we obtain $ \mbox{rank} \, | Y^ i_{(p)} g_{ij} Y^j_{(p)} | = (N-1) $, that is, the space of the orbits of this group forms an isotropic surface.

The results of Vladimir Shapovalov made it possible for the first time to carry out a complete classification of all spaces admitting the integration of the Hamilton-Jacobi equation of test particles by the method of complete separation of variables, for which in the ''privileged'' coordinate systems one of the ''separated'' nonignored variables is the ''wave'' (otherwise null or isotropic) and find an explicit form of their metrics. We will call such spaces wave-like ''Shapovalov spaces'' (more precise definitions are given below).
\begin{Opredelenie}
\label{wave_x}
We will call the coordinate variable {\bf isotropic},
if along the coordinate line of this variable the space-time interval is equal to zero.
\end{Opredelenie}
\begin{Opredelenie}
\label{Shapovalov_space}
A space-time that admits complete separation of variables in the Hamilton-Jacobi equation (\ref{HJE}) for a test particle in a gravitational field will be called {\bf Shapovalov space} if in a privileged coordinate system where separation of variables is allowed, there is a nonignored isotropic variable.
\end{Opredelenie}
\begin{Opredelenie}
\label{Conformal_Shapovalov_space}
A space-time admitting complete separation of variables in the eikonal equation (\ref{Eikonal}) will be called {\bf conformal  Shapovalov space} if there is a nonignored isotropic variable in a privileged coordinate system where separation of variables is allowed.
\end{Opredelenie}

In total, for the four-dimensional space-time, there are three main classes of Shapovalov spaces in accordance with 
the number of commuting Killing vectors in the complete set (from one to three vectors).
The form of the metrics in the privileged coordinate system is shown below (further, coordinates are numbered starting from 0).

As an example of the application of Shapovalov's wave-like spaces in metric theories of gravity, solutions of the Einstein vacuum equations with the cosmological constant $\Lambda$ are obtained:
\begin{equation}
R_{ij}-\frac{1}{2}\, R\, g_{ij}=\Lambda\, g_{ij},
\qquad
\Lambda=\mbox{const}.
\label{EinsteinEqs}
\end{equation}
The obtained solutions of the field equations for Shapovalov's wave-like spaces, when the eikonal equation and the Hamilton-Jacobi equation admit the separation of isotropic variables on which the space-time metric depends, can be interpreted as gravitational waves. Comparative analysis of exact gravitational-wave models for Shapovalov spaces in modified theories of gravity \cite{Capozziello2011167,Capozziello2008357,Nojiri201159,Nojiri2007115,Bamba2012155}) provides an additional tool for comparing and selecting viable theories of modified gravity.

Note that some of the results below could have previously been presented in some form in our other works and in the works of other researchers. The purpose of this work is to give a systematic presentation of the topic under consideration, therefore, all the necessary results that we obtained ourselves are included here.

\section{Shapovalov space of type I} 

A type I Shapovalov space admits only one Killing vector in a complete set.
The metric of the Shapovalov space of type I in the privileged coordinate system $ \{x^i \} $ can be written in the following form:
\begin{equation}
g^{ij}=
\frac{1}{\Delta}
\left(
\begin{array}{cccc}
0  &  V^{(1)} & 0  &  0 \\
V^{(1)}  & 0  &0  & 0  \\
0  & 0   &V^{(2)} &  0 \\
 0  & 0  & 0  &  V^{(3)} 
\end{array}
\right)
,
\label{metrI}
\end{equation}
$$
V^{(1)}= t_2(x^2)-t_3(x^3),
\qquad
V^{(2)}= t_3(x^3)-t_1(x^1),
\qquad
V^{(3)}= t_1(x^1)-t_2(x^2).
$$
Here, in the case of conformal Shapovalov spaces, $ \Delta $ is an arbitrary function of all variables, and for Shapovalov spaces, the conformal factor is
$\Delta=\sigma_1(x^1)V^{(1)}+\sigma_2(x^2)V^{(2)}+\sigma_3(x^3)V^{(3)}$. 
The variable $ x^0 $ is an ignored variable and $ x^1 $ is nonignored isotropic wave variable.

The space-time interval takes the form:
\begin{equation}
ds^2=\Delta\,\left(
\frac{2\,dx^0\,dx^1}{ V^{(1)}} + \frac{{dx^2}^2}{V^{(2)}} + \frac{{dx^3}^2}{V^{(3)}}
\right).
\label{metrID}
\end{equation}
The determinant of the metric is
\begin{equation}
\det{g_{ij}}=-\frac{\Delta^4}{ { V^{(1)}}^2{V^{(2)}}{V^{(3)}} }.
\end{equation}
The solution for the metric (\ref{metrID}) of the Einstein equations in vacuum (\ref{EinsteinEqs}) leads to a degeneration of the type of space: additional commuting Killing vector fields appear, the number of nonignored variables in the metric decreases and, thus, a transition to other types Shapovalov spaces  occurs (considered below). Thus, there are no solutions to the Einstein vacuum equations (\ref{EinsteinEqs}) with a cosmological constant for Shapovalov spaces of type~{I}.

Note that, in modified theories of gravity, it is possible for Shapovalov spaces of type {I} to have an exact “wave” solution of field equations in vacuum, and this problem requires additional study.

\subsection{Integration of the eikonal equation for the Shapovalov space of type~{I}}

Separation of variables in the eikonal equation (\ref{Eikonal}) for the metric (\ref{metrI}) gives the eikonal function of the form
\begin{equation}
\Psi={ \lambda_{(0)} } \, x^0 + \psi_1(x^1)+ \psi_2(x^2)+ \psi_3(x^3)+F({ \lambda_{(0)} } , { \lambda_{(1)} } , { \lambda_{(2)} }),
\label{IntegralEikonalI}
\end{equation}
$$
{ \lambda_{(0)} } , { \lambda_{(1)} } , { \lambda_{(2)} }  - \mbox{const},
$$
where $F({ \lambda_{(0)} } , { \lambda_{(1)} } , { \lambda_{(2)} })$ is an arbitrary function of constant separation parameters, and the functions $ \psi $ in (\ref{IntegralEikonalI}) are defined by the expressions
\begin{equation}
2\,{ \lambda_{(0)} }\, \psi_1={ \lambda_{(2)} }\,  x^1+{ \lambda_{(1)} } \int t_1(x^1)\,dx^1,
\end{equation}
\begin{equation}
\psi_2=\pm\int\sqrt{{ \lambda_{(1)} }  t_2(x^2) + { \lambda_{(2)} } }\,dx^2,
\qquad
\psi_3=\pm\int\sqrt{{ \lambda_{(1)} }  t_3(x^3) + { \lambda_{(2)} } }\,dx^3.
\end{equation}

\subsection{Integration of the Hamilton-Jacobi equation 
of a test particle for the Shapovalov space of type I}

Separation of variables in the Hamilton-Jacobi equation of a test particle (\ref{HJE}) in a privileged coordinate system for the metric (\ref{metrI}) gives the complete integral of the action function of a test particle of mass $ m $ in the form
\begin{equation}
S={ \lambda_{(0)} } \, x^0 + \phi_1(x^1)+ \phi_2(x^2)+ \phi_3(x^3)+F(m, { \lambda_{(0)} } , { \lambda_{(1)} } , { \lambda_{(2)} }),
\label{IntegralSI}
\end{equation}
$$
{ \lambda_{(0)} } , { \lambda_{(1)} } , { \lambda_{(2)} }  - \mbox{const},
$$
where $F(m,{ \lambda_{(0)} } , { \lambda_{(1)} } , { \lambda_{(2)} })$ is an arbitrary function of constant parameters, and the functions $ \phi $ in (\ref{IntegralSI}) are defined by the expressions
\begin{equation}
2\,{ \lambda_{(0)} }\, \phi_1={ \lambda_{(2)} }  x^1+ \int \left[m^2\sigma_1(x^1)+ { \lambda_{(1)} }\, t_1(x^1)\right]\,dx^1,
\end{equation}
\begin{equation}
\phi_2=\pm\int\sqrt{m^2\sigma_2(x^2)+{ \lambda_{(1)} }\,  t_2(x^2) + { \lambda_{(2)} } }\,dx^2,
\end{equation}
\begin{equation}
\phi_3=\pm\int\sqrt{m^2\sigma_3(x^3)+{ \lambda_{(1)} }\,  t_3(x^3) + { \lambda_{(2)} } }\,dx^3.
\end{equation}

\section{Type II Shapovalov spaces}

Type II Shapovalov spaces admit two commuting Killing vector fields in the complete set and have two subtypes: {II.A} and {II.B}. 
For all types of spaces, below, complete integrals are presented for the eikonal function and for the action function of test particles, and the Einstein equations with the cosmological constant in vacuum are integrated.

\subsection{Type II.A Shapovalov spaces}

In a privileged coordinate system, the metric can be represented as
\begin{equation}
g^{ij}=
\frac{1}{\Delta}
\left(
\begin{array}{cccc}
1   &  0 & 0  &  0 \\
0   & 0  & f_1(x^1)  & 1  \\
0  & f_1(x^1)   &a_0(x^0)+a_1(x^1) &  0 \\
 0  & 1  & 0  &  0 
\end{array}
\right)
.
\label{metrIIA}
\end{equation}
In the case of conformal Shapovalov spaces, $ \Delta $ is an arbitrary function of all variables, and in the case of Shapovalov spaces, the conformal factor is $\Delta = t_0 (x ^ 0) + t_1 (x ^ 1) $.
Variables $ x^2 $ and $ x^3 $ are ignored, $ x^1 $ is a nonignored isotropic variable.

The Shapovalov space-time interval of the type~{II.A} can then be written as:
\begin{equation}
ds^2={\Delta}\,\Bigl[
{dx^0}^2+
2\,{dx^1}{dx^3}+
\frac{1}{ {a_0}+{a_1}}\,
\left(
{dx^2}-{f_1}\,{dx^3}
\right)^2
\Bigr]
\end{equation}
The determinant of the metric {II.A} has the form
\begin{equation}
g=\det{g_{ij}}=-\frac{\Delta^4}{ { {a_0}+{a_1}} },
\qquad 
{a_0}+{a_1}>0
.
\end{equation}

\subsection{Exact solution of the Einstein equations 
for \mbox{Shapovalov} spaces of type~II.A}

For the metric (\ref{metrIIA}), we obtain the solution of the Einstein vacuum equations with
cosmological constant $ \Lambda $ (where $ x^1 $ is an isotropic variable):
\begin{equation}
{\Delta}={t_0}(x^0),
\quad
t_1=0,
\quad
a_0=1/{b_0}(x^0),
\quad
a_1=0,
\quad
{f_1}
={\alpha}\, {x^1},
\quad
{\alpha}-\mbox{const},
\end{equation}
\begin{equation}
ds^2={t_0}\left[
\,{dx^0}^2+
2\, dx^1dx^3+
{b_0}
\left(
dx^2-
\alpha\,x^1\,dx^3
\right)^2
\,
\right],
\end{equation}
\begin{equation}
{\det g_{ij}=}-{b_0}{} {t_0}{}^4,
\qquad
b_0>0.
\end{equation}
The functions $ b_0 (x^0) $ and $ t_0 (x^0) $, included in the metric, are determined through the auxiliary function $ Y (t_0) $, which is a solution to an ordinary differential equation of the second order:
\begin{equation}
3Y\frac{d^2Y}{{dt_0}^2}-\left(\frac{dY}{dt_0}\right)^2-
2\Lambda{t_0}^2\frac{dY}{dt_0}-12\Lambda{t_0}Y+8{\Lambda}^2{t_0}^4=0,
\label{Eq2IIA}
\end{equation}
where $ \Lambda $ is the cosmological constant.

Then the function $ t_0 (x^0) $ is determined by integrating the relation:
\begin{equation}
\frac{d\, t_0(x^0)}{d x^0}=\pm\sqrt{Y(t_0)}
\end{equation}
The function $ b_0 $ is defined through the function $ t_0 $ by the relation:
\begin{equation}
b_0(t_0)=
\frac{2 {t_0} Y'(t_0)-3{Y(t_0)}^2-4 \Lambda {t_0}^3}{3{\alpha}^2
   {t_0}^2}
\end{equation}
Note that the order of the differential equation (\ref{Eq2IIA}) can be lowered to the first order.

When changing variables of the form
\begin{equation}
t_0=\exp t,
\qquad
Y= X(t) \exp 3t
\end{equation}
the newly obtained differential equation for $ X (t) $ ceases to include the independent variable $ t $ and its order can be reduced by introducing a new function $ Z (X) = d X (t) / d t $.

The function $ Z (X) $, in turn, is a solution to the ordinary differential equation of the first order
\begin{equation}
3XZ(X)\frac{dZ(X)}{dX}+Z(X)\Bigl( -Z(X) +9X-2\Lambda  \Bigr)+(3X-2\Lambda)(3X-4\Lambda)=0.
\label{Eq3IIA}
\end{equation}
For the equation (\ref{Eq3IIA}), there are particular solutions of the form $ Z (X) = \beta X + 2 \Lambda $, where $ \beta = -3 $ or $ \beta = -3 / 2 $.
For these particular solutions, the function $ Y (t_0) $ takes the form
\begin{equation}
Y(t_0)={ t_0}^{3}\left(-2\Lambda/ \beta+\gamma{ t_0}^{\beta} \right),
\qquad
\gamma - \mbox{const}.
 \label{Eq3IIA1}
\end{equation}
For a particular solution (\ref{Eq3IIA1}) at $ \beta = -3 $, the function $ b_0 (x^0) $ becomes negative, which violates the requirement for the sign of the determinant of the metric. For a particular solution for $ \beta = -3 / 2 $ the function $ b_0 $ vanishes, which leads to the degeneration of the metric. However, in general, equations (\ref{Eq2IIA}), (\ref {Eq3IIA}) may have viable solutions.

Ricci tensor and scalar curvature are nonzero and proportional to the cosmological constant~$ \Lambda $. Weyl tensor and Riemann curvature tensor do not vanish.

\subsection{Integration of the eikonal equation for a Shapovalov space of type II.A}

Separation of variables in the eikonal equation for the metric of the Shapovalov space of type {II.A} gives
\begin{equation}
\Psi=\psi_0(x^0)+\psi_1(x^1)+ \lambda_{(2)} \,x^2+\lambda_{(3)} \,x^3 +F\left(\lambda_{(1)}  , \lambda_{(2)} , \lambda_{(3)}\right),
\end{equation}
\begin{equation}
\lambda_{(1)}  , \lambda_{(2)} , \lambda_{(3)}  - \mbox{const},
\label{IntegralEikonalIIA}
\end{equation}
where  $F\left(\lambda_{(1)}  , \lambda_{(2)} , \lambda_{(3)}\right)$ is an arbitrary function of the separation parameters,
\begin{equation}
\psi_0=\pm\int\sqrt{
\lambda_{(1)}-{ \lambda_{(2)} }^2\, a_0(x^0)
}\,dx^0,
\end{equation}
and the function $ \psi_1 (x^1) $ is defined by the ordinary differential equation
\begin{equation}
2\,\left[
 \lambda_{(2)}\,f_1(x^1)+ \lambda_{(3)} 
\right]\,
{\psi}_1'(x^1)
=-\lambda_{(1)}-{\lambda_{(2)} }^2\,a_1(x^1),
\end{equation}
where the prime means the ordinary derivative.

\subsection{Integration of the Hamilton-Jacobi equation of a test particle for a Shapovalov space of type II.A}

The complete integral for the action function $ S $ of a test particle of mass $ m $ takes the form
\begin{equation}
S=\phi_0(x^0)+\phi_1(x^1)+ \lambda_{(2)} \,x^2+\lambda_{(3)} \,x^3 +F\left(m,\lambda_{(1)}  , \lambda_{(2)} , \lambda_{(3)}\right),
\label{IntegralSIIA}
\end{equation}
\begin{equation}
\lambda_{(1)}  , \lambda_{(2)} , \lambda_{(3)}  - \mbox{const},
\end{equation}
where  $F\left(m,\lambda_{(1)}  , \lambda_{(2)} , \lambda_{(3)}\right)$ is an arbitrary function of the separation parameters,
\begin{equation}
\phi_0=\pm\int\sqrt{
m^2\, t_0(x^0)-{ \lambda_{(2)} }^2\, a_0(x^0)+\lambda_{(1)}
}\,dx^0,
\end{equation}
and the function $ \phi_1 (x^1) $ is determined by the ordinary differential equation
\begin{equation}
2\,\left[
 \lambda_{(2)}\,f_1(x^1)+ \lambda_{(3)} 
\right]\,
{\phi}_1'(x^1)
=m^2\,t_1(x^1)-{\lambda_{(2)} }^2\,a_1(x^1)-\lambda_{(1)}.
\end{equation}

\section{Shapovalov spaces II.B}
The metric of the Shapovalov II.B space in the privileged coordinate system can be represented as follows:
\begin{equation}
g^{ij}=
\frac{1}{\Delta}
\left(
\begin{array}{cccc}
1   &  0 & 0  &  0 \\
0   & 0  & f_1(x^1)  & 1  \\
0  & f_1(x^1)   &a_0(x^0)  b_0(x^0)  &  b_0(x^0) \\
 0  & 1  & b_0(x^0)   & b_0(x^0) / a_0(x^0)  
\end{array}
\right)
,
\label{metrIIB}
\end{equation}
moreover, $ {a_0} {b_0} \ne 0 $.
The variable $ x^1 $ is a nonignored isotropic (wave) variable.

The interval {II.B} is:
\begin{equation}
ds^2=\Delta\,\left[
{dx^0}^2
+\frac{2\, dx^1\left(
- dx^2+a_0\,dx^3
\right)}{ a_0-f_1}+
\frac{a_0 \left(
dx^2-f_1\,dx^3
\right)^2}{ b_0\,( a_0-f_1)^2}
\right].
\end{equation}
The determinant of the metric {II.B} has the form
\begin{equation}
g=\det{g_{ij}}=-\frac{a_0\,\Delta^4}{b_0\,(a_0-f_1)^2 },
\qquad
a_0\,b_0>0.
\end{equation}
When integrating field equations for Shapovalov spaces, functional equations often arise that connect the functions of the metric in different variables.

The Einstein equations (\ref{EinsteinEqs}) with indices of nonignored variables $ \{0,1 \} $ can be written for the metric {II.B} in the form
\begin{equation}
\partial_0\partial_1\left( \ln\left[\frac{\Delta}{a_0(x^0)-f_1(x^1)}\right]    \right)=0.
\end{equation}
For the case of St{\"{a}}ckel spaces, when $ \Delta = {t_0 (x^0) + t_1 (x^1)} $, we obtain the functional equation:
\begin{equation}
\partial_0\partial_1\left( \ln\left[\frac{t_0(x^0)+t_1(x^1)}{a_0(x^0)-f_1(x^1)}\right]    \right)=0.
\label{eq01}
\end{equation}
The equation (\ref{eq01}), considered as a functional equation, taking into account the condition $ {f_1}' \ne 0 $ (otherwise the degeneration of the separation type occurs), has the following set of solutions:
\begin{eqnarray}
1.&\Delta=&\frac{1}{f_1(x^1)+q}-\frac{1}{a_0(x^0)+q},  \quad q=\mbox{const}; \label{case1} \\
2.&\Delta=&a_0(x^0)-f_1(x^1);\\
3.&\Delta=&t_0(x^0), \quad t_1=0, \quad a_0=\mbox{const};\\
4.&\Delta=&t_1(x^1), \quad t_0=0, \quad a_0=\mbox{const}.
\end{eqnarray}
Note that cases 2, 3 and 4 when solving the Einstein vacuum equations with a cosmological constant (\ref{EinsteinEqs}) lead to contradictions, and only in case 1 we obtain two exact solutions of the field equations, which are listed below.

\subsection{Exact solution \#1 of Einstein's equations for II.B type Shapovalov space}

The first exact solution of the Einstein vacuum equations
for a metric (\ref{metrIIB}) of type {II.B} in case (\ref{case1}) has the form (where $ x^1 $ is an isotropic wave variable):
$$
a_0={a_0(x^0)},
\qquad
b_0={b_0(x^0)},
\qquad
f_1={f_1(x^1)},
$$
$$
ds^2=\frac{1}{ ({a_0}+q)({f_1}+q)}\,\Bigl[
({f_1}-{a_0})\,{dx^0}^2
+2\,dx^1\left(
dx^2
-a_0\,dx^3
\right)
$$
\begin{equation}
\mbox{}+
\frac{{a_0}}{{b_0}({f_1}-{a_0})}\,
\left(
{dx^2}-{f_1}\,{dx^3}
\right)^2
\Bigr].
\end{equation}
The cosmological constant vanishes, the solution has six independent constants:
\begin{equation}
\Lambda=0,
\qquad
{p},{q},{r},{k},{\alpha},{\beta}, {\gamma} - \mbox{const},
\qquad
\gamma=-{q}\pm\, \sqrt{\alpha^2+\beta^2},
\end{equation}
Functions included in the conformal factor of the metric,
take the form:
\begin{equation}
t_0=\frac{1-{p}{x^0}^4}{2 ({q}+\gamma )}+{r},
\qquad
t_1=-\frac{1}{{f_1(x^1)}+{q}}-{r},
\end{equation}
\begin{equation}
\Delta=
-\frac{\left({p}{x^0}^4-1\right) {f_1(x^1)}{}+{p}{q} {x^0}^4+{q}+2 \gamma }{2 ({q}+\gamma )
   ({f_1(x^1)}{}+{q})}.
\end{equation}
The functions $ a_0 (x^0) $, $ b_0 (x^0) $ and $ f_1 (x^1) $, included in the metric, are defined by the relations:
\begin{equation}
f_1(x^1)=
\alpha  \cos ({k} {x^1})+\beta  \sin ({k} {x^1})+\gamma,
\end{equation}
\begin{equation}
{a_0}(x^0)=
\frac{{q} \left( {p} {x^0}^4+1\right)+2 \gamma }{1-{p}
   {x^0}^4},
\end{equation}
\begin{equation}
\Delta\,g^{23}=
{b_0}(x^0)=
-\frac{{k}^2 \left({p}{x^0}^4-1\right)
   \left({q} \left( {p} {x^0}^4+1\right)+2 \gamma \right)}{16
   {p}{x^0}^2},
\end{equation}
\begin{equation}
\Delta\,g^{22}=
{a_0}\,{b_0}=
\frac{{k}^2 \left({q} \left( {p} {x^0}^4+1\right)+2 \gamma \right)^2}{16 {p}{x^0}^2},
\end{equation}
\begin{equation}
\Delta\,g^{33}=
{b_0}/{a_0}=
\frac{{k}^2 \left({p}{x^0}^4-1\right)^2}{16 {p}{x^0}^2},
\end{equation}
The obtained metric depends on the wave variable~$ x^1 $  through harmonic functions.

The determinant of the metric is:
\begin{equation}
{\det g_{ij}=}-\frac{{p}\, {x^0}^2 \left(\left({p}{x^0}^4-1\right) {f_1(x^1)}{}+{q} \left( {p} {x^0}^4+1\right)+2 \gamma \right)^2}{{k}^2 ({q}+\gamma )^4 ({f_1(x^1)}{}+{q})^4},
\qquad
{p}>0,
\end{equation}
For this solution, the Ricci tensor $ R_{ij} $, the scalar curvature $ R $ and the cosmological constant $ \Lambda $ vanish. The Riemann curvature tensor and the Weyl tensor do not vanish.

\subsection{Exact solution \#2 of Einstein's equations for Shapovalov space of type II.B}

The second exact solution of Einstein's vacuum equations for metric (\ref{metrIIB}) in case (\ref{case1})
has four independent constant parameters, the cosmological constant vanishes:
\begin{equation}
\Lambda=0,
 \qquad
{p},{q},{r},{k} - \mbox{const},
\qquad
{p}\,{q}<0,
\end{equation}
Apart from the conformal factor $ \Delta (x^0, x^1) $, the metric is determined by three functions (here the variable $ x^1 $ is an ignored isotropic variable):
$$
a_0={a_0(x^0)},
\qquad
b_0={b_0(x^0)},
\qquad
f_1={f_1(x^1)}.
$$
The space-time interval takes the form:
$$
ds^2=\frac{1}{ ({a_0}+q)({f_1}+q)}\,\Bigl[
( {f_1}-{a_0})\,{dx^0}^2
+2\,dx^1\left(
dx^2
-a_0\,dx^3
\right)
$$
\begin{equation}
\mbox{}+
\frac{{a_0}}{{b_0}\,( {f_1}-{a_0})}
\left(
{dx^2}-{f_1}\,{dx^3}
\right)^2
\Bigr].
\end{equation}
The functions included in the conformal factor of the metric $ \Delta $ are determined by the expressions:
\begin{equation}
t_0(x^0)=
\frac{{x^0}^4}{{p}}-{k},
\qquad
t_1(x^1)=
{k}-\frac{1}{{q}\, {x^1}^2},
\qquad
\Delta(x^0,x^1)=
\frac{{x^0}^4}{{p}}-\frac{1}{{q} {x^1}^2},
\end{equation}
The functions $ a_0 (x^0) $, $ b_0 (x^0) $ and $ f_1 (x^1) $ and the components of the metric are determined by the relations:
\begin{equation}
{f_1}(x^1)=
{q} \, {x^1}^2-{r},
\end{equation}
\begin{equation}
{a_0}(x^0)=
\frac{{p}}{{x^0}^4}-{r},
\end{equation}
\begin{equation}
\Delta\,g^{23}={b_0}(x^0)=
\frac{{q}}{4{p}}\, {x^0}^2 \left({r} {x^0}^4-{p}\right),
\end{equation}
\begin{equation}
\Delta\,g^{22}={a_0}(x^0)\,{b_0}(x^0)=
-\frac{{q} \left({p}-{r} {x^0}^4\right)^2}{4 {p} {x^0}^2},
\end{equation}
\begin{equation}
\Delta\,g^{33}={a_0}(x^0)/{b_0}(x^0)=
-\frac{4 {p}}{{q} {x^0}^6},
\end{equation}
\begin{equation}
{f_1}(x^1)-{a_0}(x^0)=
{q} {x^1}^2-\frac{{p}}{{x^0}^4}.
\end{equation}
The determinant of the metric takes the form:
\begin{equation}
{g=\det g_{ij}=}\frac{4 {x^0}^2 \left({p}-{q} {x^0}^4 {x^1}^2\right)^2}{{p}^3 {q}^5 {x^1}^8},
\qquad
{p}\,{q}<0,
\end{equation}
For a given exact solution, the Ricci tensor, scalar curvature and cosmological constant vanish. The Riemann curvature tensor and the Weyl tensor do not vanish.

\subsection{Integration of the eikonal equation for the Shapovalov space of type II.B}

Separation of variables in the eikonal equation (\ref{Eikonal}) for a Shapovalov space of type {II.B} in a privileged coordinate system with metric (\ref{metrIIB}) gives the following form of the eikonal function
\begin{equation}
\Psi=\psi_0(x^0)+\psi_1(x^1)+\lambda_{(2)}\,x^2+\lambda_{(3)}\,x^3,
\qquad
\lambda_{(1)}  , \lambda_{(2)} , \lambda_{(3)} - \mbox{const},
\end{equation}
where the functions $ \psi_0 $ and $ \psi_0 $ are defined by the relations
\begin{equation}
\psi_0=\pm\int{
\sqrt{
\lambda_{(1)} -\frac{a_0(x^0)}{b_0(x^0)}
\Bigl(
\lambda_{(2)}\,a_0(x^0)+\lambda_{(3)} 
\Bigr)^2
}
}\,dx^0,
\end{equation}
\begin{equation}
\psi_1=-\frac{\lambda_{(1)} }{2}
\int{
\frac{
dx^1
}{
\lambda_{(2)}\,f_1(x^1)+ \lambda_{(3)}
}
}.
\end{equation}

\subsection{Integration of the Hamilton-Jacobi equation of motion of a test particle for the Shapovalov space of type II.B}

Separation of variables in the Hamilton-Jacobi equation (\ref{HJE}) for Shapovalov spaces of type {II.B} with metric (\ref{metrIIB}) in a privileged coordinate system gives the full integral of the action function of a test particle of mass $ m $ of the form:
\begin{equation}
\Psi=\phi_0(x^0)+\phi_1(x^1)+\lambda_{(2)}\,x^2+\lambda_{(3)}\,x^3,
\qquad
\lambda_{(1)}  , \lambda_{(2)} , \lambda_{(3)} - \mbox{const},
\end{equation}
where the functions $ \phi_0 $ and $ \phi_1 $ are defined by the following relations:
\begin{equation}
\phi_0=\pm\int{
\sqrt{
m^2\,t_0(x^0) +    \lambda_{(1)} -\frac{a_0(x^0)}{b_0(x^0)}\,
\Bigl(
\lambda_{(2)}\,a_0(x^0)+\lambda_{(3)} 
\Bigr)^2
}
}\,dx^0,
\end{equation}
\begin{equation}
\phi_1=\frac{1}{2}
\int{
\frac{
m^2\,t_1(x^1)  -  \lambda_{(1)} 
}{
\lambda_{(2)}\,f_1(x^1)+ \lambda_{(3)}
}
}\,dx^1.
\end{equation}
In case when the function $ f_1 $ is a constant, it can be set equal to zero by transformation of coordinates.

\section{Type III Shapovalov spaces}

A Type III Shapovalov space admits three commuting Killing vectors in a complete set.
Metric of a Shapovalov space of type III in the privileged coordinate system
can be written as follows:
\begin{equation}
g^{ij}=
\frac{1}{\Delta}
\left(
\begin{array}{cccc}
0                   &  1             & g^{02}(x^0)  &  g^{03}(x^0) \\
1                   & 0  & 0  & 0  \\
g^{02}(x^0)  & 0   &g^{22}(x^0)  &  g^{23}(x^0) \\
g^{03}(x^0)  & 0  & g^{23}(x^0)  & g^{33}(x^0)
\end{array}
\right)
.
\label{MetrIII}
\end{equation}
In the case of Shapovalov spaces, the conformal factor can be set equal to one ($ \Delta = 1 $), and in the case of conformal Shapovalov spaces, the conformal factor $ \Delta $ is an arbitrary function of all four variables.

The metric interval (\ref{MetrIII}) can be written in the following form ($p,q=2,3$):
\begin{equation}
ds^2=\Delta\left[
2\,dx^0dx^1+g_{pq}(x^0)\left(dx^p+g^{(p)}(x^0)\,dx^1\right)\left(dx^q+g^{(q)}(x^0)\,dx^1\right)
\right].
\label{MetrIIID}
\end{equation}
The last expression, disregarding the conformal factor $ \Delta $, contains five arbitrary functions of one variable $ x^0 $, as in the metric (\ref{MetrIII}). Note that the metric (\ref{MetrIIID}) is known in the literature as the metric of a gravitational wave (see \cite{LandauEng2}).

\subsection{Exact solution of the Einstein equations for  \mbox{Shapovalov} spaces type III }

Integration of Einstein's equations in vacuum for the metric (\ref{MetrIIID}) leads to an exact solution (gravitational wave), which can be represented as
$$
ds^2=2\,dx^0dx^1- \exp{(-\gamma_0)}
\Bigl(
\exp{\beta_0}\cosh{\alpha_0}\,\,{dx^2}^2+
\exp{(-\beta_0)}\cosh{\alpha_0}\,\,{dx^3}^2
$$
\begin{equation}
\mbox{}+
2\,\sinh{\alpha_0}\,\,dx^2dx^3
\Bigr),
\qquad
\Lambda=0,
\end{equation}
and only two functions are independent, the function $ \beta_0 $ can be expressed in terms of the rest
\begin{equation}
\beta_0=\int{
\frac{
\sqrt{2\,\gamma_0''-{ \gamma_0'}{}^2-{\alpha_0'}^2\,
}
}{ \cosh{\alpha_0} }
}\,dx^0,
\qquad
\Lambda=0,
\end{equation}
where $ \alpha_0 $ and $ \gamma_0 $ are arbitrary functions of one variable $ x^0 $ (wave variable), and the prime means ordinary differentiation with respect to $ x^0 $.

\subsection{Integration of the eikonal equation for the \mbox{Shapovalov} space type III}

Separation of variables in the eikonal equation (\ref{Eikonal}) for the metric (\ref{MetrIII}) gives the complete integral for the eikonal function of the form
\begin{equation}
\Psi=\psi_0(x^0)+\lambda_{(1)} \, x^1 + \lambda_{(2)} \,x^2+\lambda_{(3)} \,x^3 +F\left(\lambda_{(1)}  , \lambda_{(2)} , \lambda_{(3)}\right),
\label{IntegralEikonalIII}
\end{equation}
$$
\lambda_{(1)}  , \lambda_{(2)} , \lambda_{(3)}  - \mbox{const},
$$
where  $F\left(\lambda_{(1)}  , \lambda_{(2)} , \lambda_{(3)}\right)$ is an arbitrary function of parameters, and the function $ \psi_0 (x^0) $ in (\ref{IntegralEikonalIII}) is defined by the following expression
\begin{equation}
{\psi_0}(x^0)=-\frac{1}2{}
\int{
\frac{
\lambda_{(p)}\lambda_{(q)}g^{pq}(x^0)
}{
\lambda_{(1)} +\lambda_{(p)}\,g^{0p}(x^0)
}
}\, dx^0,
\qquad
p,q=2,3.
\end{equation}
When the components of the metric $ g^{0p} $ become constants, they can be converted to zero by transforming the coordinates.

\subsection{Integration of the Hamilton-Jacobi equation of a test particle for Shapovalov space type~III}

Separation of variables in the Hamilton-Jacobi equation of a test particle (\ref{HJE}) for the metric (\ref{MetrIII}) gives a complete integral for the action function of a test particle of mass $ m $ of the form
\begin{equation}
S=\phi_0(x^0)+\lambda_{(1)} \, x^1 + \lambda_{(2)} \,x^2+\lambda_{(3)} \,x^3 +F\left(m,\lambda_{(1)}  , \lambda_{(2)} , \lambda_{(3)}\right),
\label{IntegralSIII}
\end{equation}
$$
\lambda_{(1)}  , \lambda_{(2)} , \lambda_{(3)}  - \mbox{const},
$$
where $F\left(m,\lambda_{(1)}  , \lambda_{(2)} , \lambda_{(3)}\right)$ is an arbitrary function of parameters, and the function $\phi_0(x^0)$ в (\ref{IntegralSIII}) 
is defined by the following expression
\begin{equation}
\phi_0(x^0)=
\frac{1}{2}
\int \frac{m^2-\lambda_{(p)}\lambda_{(q)}g^{pq}(x^0)}{
\lambda_{(1)} +\lambda_{(p)}\,g^{0p}(x^0)
}\,dx^0,
\qquad
p,q=2,3.
\end{equation}
Note that the variables $ x^0 $ and $ x^1 $ of the privileged coordinate system are isotropic variables,
and the space-time metric depends only on one wave variable $ x^0 $ and is interpreted as an exact solution for a gravitational wave.

\section*{Conclusion}

The paper presents a mathematical tool for studying wave models in the theory of gravity and cosmology, based on the classification of space-time models that allow the integration of the eikonal equation and the Hamilton-Jacobi equation for test particles in a gravitational field by the method of complete separation of variables with the separation of isotropic (wave) of nonignored variables. The considered space-time models admit the 
privileged coordinate systems, where it is possible to separate wave variables in the equations of motion of test massless and massive particles, and the metric depends on these wave variables. We call such spaces wave-like Shapovalov spaces. The classification of Shapovalov spaces is based on the number of commuting Killing vector fields admitted by these spaces (the dimension of the Abelian subgroup of motions of the space). An explicit form of the metric tensor in a privileged coordinate system for all types of considered spaces is presented. The results of integration of the eikonal equation and the Hamilton-Jacobi equation for test particles are presented. For the spaces under consideration exact solutions of
Einstein's equations with a cosmological constant in a vacuum are obtained. The resulting exact solutions describe
gravitational waves in a vacuum and can be used for comparison with similar solutions in modified theories of gravity.

\section*{Acknowledgments}

The reported study was funded by RFBR, project number N~20-01-00389~A.



%

\end{document}